\documentclass[conference,a4paper]{./IEEEtran}
\usepackage{xcolor}

\usepackage{cite}

\ifCLASSINFOpdf
  \usepackage[pdftex]{graphicx}
\else
  \usepackage[dvips]{graphicx}
\fi
\usepackage{amsmath}
\usepackage{amsfonts,amssymb,amsthm}
\usepackage{upgreek}
\usepackage[tight]{subfigure}
\usepackage{multirow}
\usepackage{tabularx}
\usepackage[margin=0.9in]{geometry}

\DeclareRobustCommand*{\IEEEauthorrefmark}[1]{\raisebox{0pt}[0pt][0pt]{\textsuperscript{\footnotesize #1}}}

\begin{document}
%
\title{Terahertz Dielectric Resonator Antenna Coupled to Graphene Plasmonic Dipole}
%
%
%

\author{\IEEEauthorblockN{
Seyed Ehsan Hosseininejad\IEEEauthorrefmark{1}, Mohammad Neshat\IEEEauthorrefmark{1}, Reza Faraji-Dana\IEEEauthorrefmark{1}, 
Sergi Abadal\IEEEauthorrefmark{2}, Max C. Lemme\IEEEauthorrefmark{4}, \\Peter Haring Bol\'{i}var\IEEEauthorrefmark{3}, Eduard Alarc\'{o}n\IEEEauthorrefmark{2}, and Albert Cabellos-Aparicio\IEEEauthorrefmark{2}
}                                     
\IEEEauthorblockA{\IEEEauthorrefmark{1}
School of Electrical and Computer Engineering University of Tehran, Tehran, Iran, sehosseininejad@ut.ac.ir}
\IEEEauthorblockA{\IEEEauthorrefmark{2}
NaNoNetworking Center in Catalonia (N3Cat), Universitat Polit\`{e}cnica de Catalunya, 08034 Barcelona, Spain}
\IEEEauthorblockA{\IEEEauthorrefmark{3}
Institute for High Frequency Electronic and Quantum Electronics, University of Siegen, Siegen, Germany}  
\IEEEauthorblockA{\IEEEauthorrefmark{4}
Faculty of Electrical Engineering and Information Technology, RWTH Aachen University, Aachen, Germany}  
}

\maketitle

\begin{abstract}
This paper presents an efficient approach for exciting a dielectric resonator antenna (DRA) in the terahertz frequencies by means of a graphene plasmonic dipole. Design and analysis are performed in two steps. First, the propagation properties of hybrid plasmonic one-dimensional and two-dimensional structures are obtained by using transfer matrix theory and the finite-element method. The coupling amount between the plasmonic graphene mode and the dielectric wave mode is explored based on different parameters. These results, together with DRA and plasmonic antenna theory, are then used to design a DRA antenna that supports the ${TE}^y_{112}$ mode at 2.4 THz and achieves a gain (IEEE) of up to 7 dBi and a radiation efficiency of up 70\%. This gain is 6.5 dB higher than that of the graphene dipole alone and achieved with a moderate area overhead, demonstrating the value of the proposed structure.
\end{abstract}



%
\IEEEpeerreviewmaketitle

\vspace{7pt}
\section{Introduction}
\label{sec:intro}
Terahertz band (0.1-10 THz) wireless communications are expected to become a key technology to not only meet the increasing data rate requirements at the macroscale \cite{Kurner2014}, but also to enable new applications at the micro/nanoscale such as the Internet of NanoThings \cite{Akyildiz2014} or software-defined metamaterials \cite{AbadalACCESS}. This new scenario will impose very stringent resource constraints and adverse channel conditions, among other challenges, which call for antennas and transceivers with unprecedented compactness and reconfigurability.


Graphene has been in the spotlight for the past decade due to its outstanding properties \cite{CastroNeto2009} and has shown promise for the implementation of terahertz antennas \cite{Correas2017}. Thanks to its support for surface plasmon polaritons (SPPs) in the terahertz band, graphene antennas can be effectively miniaturized \cite{Llatser2012}. Another outstanding property of graphene-based antennas is tunability, which allows changing the frequency of resonance by simply varying a biasing voltage \cite{Hosseininejad2016a}. Both miniaturization and tunability, however, come at the cost of relatively low efficiency due to the significant losses of the SPP modes in graphene. 


A way to address this issue is through the design of dielectric resonator antennas (DRAs). DRAs are a common technique in antenna design and is widely used in mmWave systems because of their small size, high radiation efficiency and gain, and low cost \cite{Petosa2011, Hou2014}. Here, we propose to apply methods associated to DRAs to improve the performance of plasmonic graphene terahertz antennas in an area-efficient manner. One could, for instance, excite higher-order modes to increase the antenna gain as done in DRAs at mm-wave frequencies \cite{Hou2014}.

Hybrid plasmonic structures have been introduced to provide a better balance between mode confinement and propagation loss in waveguides operating in the visible regime with metals \cite{TalafiNoghani2013} or in the terahertz band with graphene \cite{Hosseininejad2016}. The hybrid structure, at its most basic form, is composed of a layer of low index dielectric (L-layer), a layer of high index dielectric (H-layer), and the plasmonic material. This concept can be extrapolated to the design of antennas combining the miniaturization of plasmonics with the qualities of DRAs, but such possibility has been scarcely investigated in the literature. In \cite{kianinejad2016}, spoof plasmons excited by a meander slot are fed to a DRA at microwave frequencies. In \cite{malheiros2015}, a metallic coplanar waveguide couples plasmons to a DRA at optical frequencies. 

\begin{figure}[!t] 
\centering
\includegraphics[width=0.9\columnwidth]{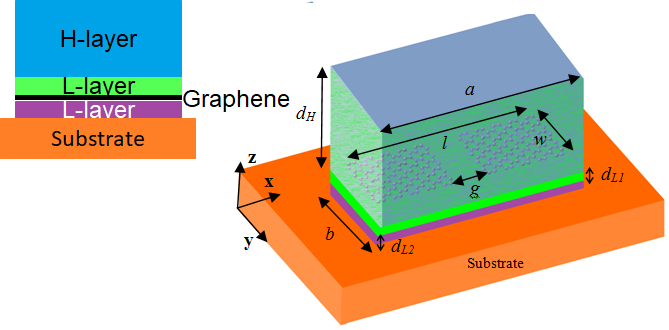}
\vspace{-0.2cm}
\caption{Configuration of a dielectric resonator antenna coupled to a graphene plasmonic dipole.}
\label{fig:structures}
\vspace{-0.3cm}
\end{figure}

This paper presents, for the first time, a THz antenna design coupling a DRA structure to a graphene dipole using the hybrid concept mentioned above. The antenna, depicted in Fig. \ref{fig:structures}, consists of a dipole of total length $l$ and width $w$. The dipole arms are composed of five-layer graphene deposited on top of a thin L-layer, polymethylmethacrylate (PMMA) in our case. The substrate is high-resistivity Gallium Arsenide (GaAs). The dipole is assumed to be fed at its center gap of length $g$ with a photoconductive source \cite{Cabellos-Aparicio2015} and placed below a $a \times b \times d_H$ rectangular resonator. The resonator is composed by a low-loss dielectric material with high permittivity (here GaAs) over a dielectric spacer (here PMMA). 

In order to design an efficient antenna, a comprehensive analysis of one-dimensional and two-dimensional hybrid plasmonic structures is performed as a basic step. Section \ref{sec:1D-2D} details the results of such analysis. Then, the dielectric resonator is designed to operate at the ${TE}_y^{111}$ and ${TE}_y^{112}$ modes when coupled to the graphene dipole based on the results of 1D-2D analysis. Section \ref{sec:3D} details the design process and simulates the resulting antenna, showing that high gain radiation efficiency is achieved by exciting the DRA structure at higher-order modes. Section \ref{sec:conc} concludes the paper.




\vspace{7pt}
\section{One-dimensional and two-dimensional hybrid plasmonic structures based on graphene}
\label{sec:1D-2D}
The initial step for designing a novel efficient antenna is studying the guided-wave structure related to it. For example, consider a patch antenna at microwave frequencies. In this case, knowing the propagation properties of a microstrip waveguide is very helpful to design the patch antenna appropriately. Similarly, here we first analyze one-dimensional hybrid plasmonic structures as a starting point for the design of realizable two-dimensional structures.

The proposed hybrid structure, which combines the plasmonic mode and the dielectric mode, allows to exploit the benefits of both guiding mechanisms. We consider few-layer graphene with five layers as a plasmonic material, PMMA with $\epsilon_{r}$ = 2.4 and thickness $d_L$ as the L-layer and GaAs $\epsilon_{r}$ = 12.9 and thickness $d_H$ as the H-layer. Therefore, the 1D structure is PMMA-Graphene-PMMA-GaAs-air. Note that the absorption losses of GaAs and PMMA are not included in the following analysis since they are small as opposed to that of graphene in the THz frequencies \cite{Zhou2014}. Wave is assumed to propagate in the x-direction and to be invariant in the y-direction. 

Few-layer graphene, instead of monolayer graphene, is adopted to construct the hybrid structure due to the following. Few-layer graphene has a lower propagation loss, which leads to a higher radiation efficiency. Besides, it provides a better coupling between the plasmonic mode and the dielectric mode because of the decreasing effective index of the plasmonic mode, which comes close to the index of the dielectric mode.

The approaches can be consider to perform the electromagnetic modeling of graphene. On the one hand, graphene can be modeled as a boundary condition that includes the complex surface conductivity $\sigma_G$ or the equivalent surface impedance $Z_G$, also complex \cite{hanson2008dyadic}. On the other hand, graphene can be represented as a layer of bulk material with small thickness $d_G$ and equivalent complex relative permittivity ($\epsilon_G=1-j\tfrac{\sigma_G}{\omega\epsilon_0d_G}$) \cite{Vakil2011}. This formalism assumes time harmonic dependence as $e^{+j{\omega}t}$. The results of the two modeling approaches are very similar for the conditions assumed in this work.

The conductivity of FLG is $N\sigma_G$ \cite{casiraghi2007rayleigh, Zhou2014}, where $N$ is the number of layers ($N<6$) and $\sigma_G$ is the conductivity of monolayer graphene which can be calculated by the well-known Kubo formula \cite{Gusynin2007} as
\begin{equation}
\begin{split}
\sigma_{G} = \frac{-j}{\omega - j \tau^{-1}} \frac{e^{2}k_{B}T}{\pi \hbar^{2}} \left (\frac{\mu_{c}}{k_{B}T} + 2ln(e^{-\frac{\mu_{c}}{k_{B}T}}+1) 
\right ) + \\
+ \frac{-j(\omega - j\tau^{-1})e^{2}}{\pi \hbar^{2}} \int_{0}^{\infty}\frac{f(-\varepsilon)+f(+\varepsilon)}{(\omega -j\tau^{-1})^{2} - 4(\varepsilon/\hbar)^{2}} d\varepsilon ,
\end{split}
\end{equation}
where $\omega$ is the radian frequency, $e$ is the electron charge, $\hbar$ is the reduced Plank constant, $k_{B}$ is the Boltzmann constant, $T$ is the temperature ($T=300$ K in this paper), $\mu_{c}$ is the chemical potential, and $\tau$ is the electron relaxation time of graphene. We consider $\tau=0.6$ ps, which is reasonable considering values in related work \cite{Correas2017}. Finally, $f(\varepsilon)=1/\{1+exp[(\varepsilon-\mu_{c})/(k_{B}T)]\}$ is the Fermi-Dirac distribution function.
In order to calculate the complex effective index of guided modes in the graphene-integrated structures, the formulations of transfer matrix theory provided in the previous work \cite{Hosseininejad2015} are applied.

\begin{figure}[!t] 
\centering
\includegraphics[width=0.9\columnwidth]{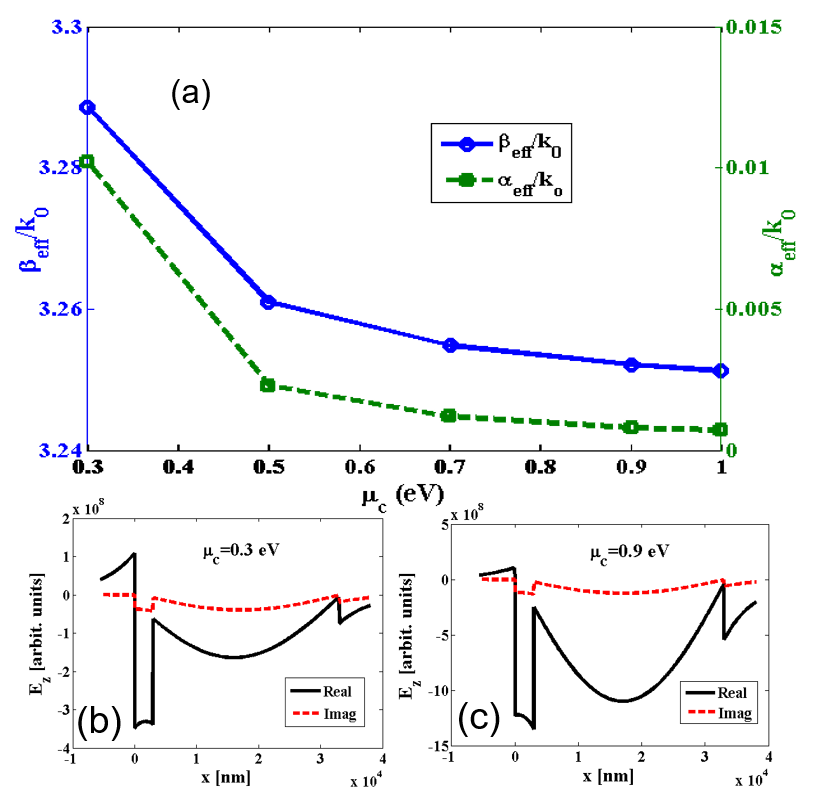} 
\vspace{-0.2cm}
\caption{Normalized propagation constant and attenuation constant hybrid 1D structure for $d_H=30$ $\upmu$m and $d_L=3$ $\upmu$m (top) and normal electric fields for $\mu_c=\{0.3, 0.9\}$ eV at 3 THz (bottom left, right).}
\label{fig:beta-mu}
\vspace{-0.3cm}
\end{figure}  

\begin{figure}[!t] 
\centering
\includegraphics[width=0.9\columnwidth]{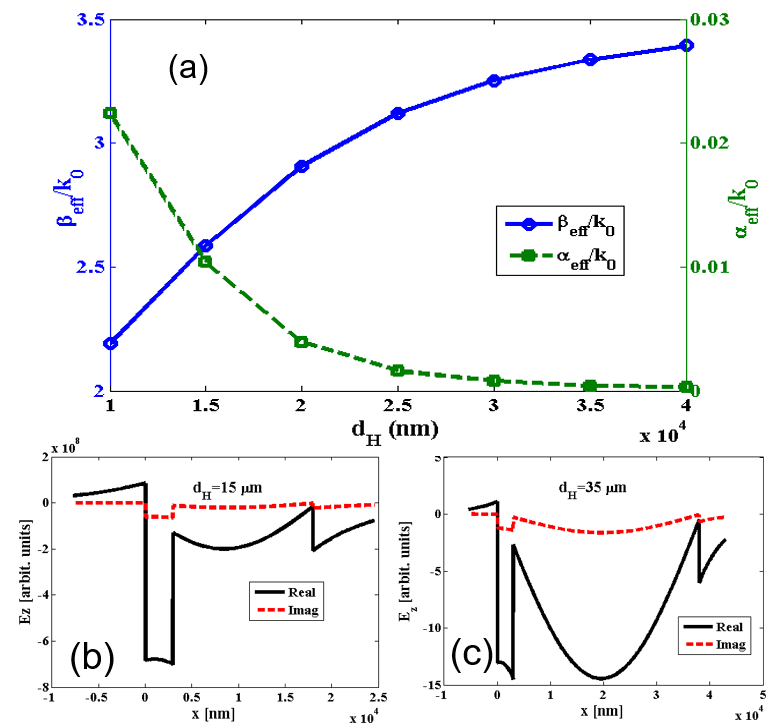}
\vspace{-0.2cm}
\caption{Normalized propagation constant and attenuation constant of the hybrid 1D structure for ${\mu}_c=0.9$ eV and $d_L=3$ $\upmu$m (top) and normal electric fields for $d_H=\{15, 35\}\,\upmu$m at 3 THz (bottom left, right).}
\label{fig:beta-dh}
\vspace{-0.3cm}
\end{figure}

\vspace{0.1cm}
\noindent
\textbf{1D analysis:} To illustrate the amount of coupling between modes, Fig. \ref{fig:beta-mu}(a) plots the normalized propagation constant $\beta_{eff}/k_0$ and normalized attenuation constant $\alpha_{eff}/k_0$ of the 1D structure for fixed dielectric thicknesses ($d_H = 30 \upmu$m, $d_L = 3 \upmu$m) and variable chemical potential. Fig. \ref{fig:beta-dh}(a) plots the same metrics for fixed chemical potential (${\mu}_c=0.9$ eV) and variable H-layer thickness with $d_L=3 \upmu$m. The frequency is 3 THz in both cases. 

Through both plots, it is observed that the chemical potential and the H-layer thickness control the coupling strength between the plasmonic and dielectric wave modes. At small $d_H$ (see Fig. \ref{fig:beta-mu}(b)) or at low ${\mu}_c$ (Fig. \ref{fig:beta-dh}(b)), the resulting hybrid mode is more \emph{SPP-like}, since most of the wave energy is concentrated at the graphene--L-layer interface, resulting in higher loss and weaker coupling as compared to structures with large $d_H$ (Fig. \ref{fig:beta-mu}(c)) or high $\mu_c$ (Fig. \ref{fig:beta-dh}(c)). Therefore, we choose sufficiently large $d_H$ and $\mu_c$ in the next design steps in order to achieve a good SPP-dielectric coupling and low loss. This design decision will lead to a more efficient antenna.

\begin{figure}[!t] 
\centering
\includegraphics[width=0.6\columnwidth]{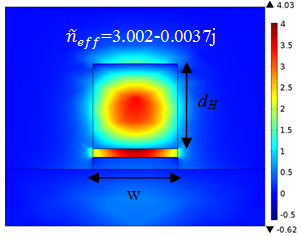}
\vspace{-0.2cm}
\caption{Normal electric field and complex effective index of hybrid two-dimensional structure for $\mu_c=0.8$ eV, $d_H=30$ $\upmu$m, $d_{L1}=3$ $\upmu$m, and $d_{L2}=4$ $\upmu$m at 3 THz.}
\label{fig:field2D}
\vspace{-0.2cm}
\end{figure}

\vspace{0.1cm}
\noindent
\textbf{2D analysis:} Let us assume a fairly high values of chemical potential $\mu_c=0.8$ eV and H-layer thickness $d_H=30$ $\upmu$m. Also, $d_{L1}=3 \upmu$m, $d_{L2}=4 \upmu$m, and $w=30 \upmu$m. We use the finite element method in COMSOL to obtain the normal electric field and complex effective index of the 2D structure. Fig. \ref{fig:field2D} shows the results at 3 THz. Good coupling is observed between the plasmonic mode and the dielectric mode as strong energy levels are confined in both in the L-layer region (SPP mode) and the H-layer region (dielectric mode). It should be noted that, in this structure, we investigate the single-mode properties because the width $w$ and height $d_H$ of the H-layer are both smaller than half of the wavelength. However, if we increase these dimensions sufficiently, high-order modes would appear. This feature is subsequently used in the design of the antenna structure in order to excite the high-order resonating modes.

\vspace{7pt}
\section{Antenna design and evaluation} 
\label{sec:3D}
As discussed in Sec. \ref{sec:1D-2D}, the graphene SPP mode can be efficiently coupled to the dielectric wave mode. This idea is applied for exciting a DRA by a graphene plasmonic dipole. 
	
Considering a rectangular DRA with relative permittivity $\varepsilon_r=12.9$ and with dimensions of $a \times b \times {d_H}$, the resonance frequency $f_0$ of the ${TE}_y^{mnp}$ mode for dielectric can be estimated from the transcendental equation as \cite{Maity2014}
\begin{equation} \label{eq:modes}
{k_y}\tan (\frac{{{k_y}b}}{2}) = \sqrt {({\varepsilon_r} - 1)k_0^2 - k_y^2} , 
\end{equation}
where  $k_x^2 + k_y^2 + k_z^2 = {\varepsilon_r}k_0^2$, ${k_x} = m\pi /a$,${k_z} = p\pi /{d_H}$, and ${k_0} = 2\pi /{\lambda _0}$. It should be mentioned that the results from Eq. \eqref{eq:modes} are related to a dielectric resonator isolated in free space and it will be an approximate value for the resonance frequency in the proposed structure \cite{Maity2014}.

On the other hand, according to transmission line model for the plasmonic transverse magnetic (TM) mode on a graphene strip, the resonance frequency of a graphene dipole can be approximately predicted as
\begin{equation} \label{eq:spp}
l = k\frac{{{\lambda _{spp}}}}{2}\,,\,\,\,\,\,\,\,\,\,\,k = 1,2,\,...
\end{equation}
where $\lambda_{spp}$ is the SPP wavelength and $k$ refers to $k$-th resonance of the dipole antenna. It is worth to note that the resonance frequency from above relation will be an approximate value because of the non-negligible capacitance of the dipole gap and also the presence of fringing fields in this geometry \cite{tamagnone2014}. The first resonance ($k=1$) of any dipole antenna has a low input impedance, and it may be seen as a short-circuit resonance (imaginary part is zero). Its second resonance ($k=2$), where the real and imaginary parts of input impedance are large, can be interpreted as an open-circuit (OC) resonance. Contrary to in RF/microwave antennas, the open-circuit resonance is desired for plasmonic THz antennas because of the decreasing return loss obtained when connecting the antenna to photomixer THz sources that typically have very high impedance.

\begin{figure}[!t] 
\centering
\includegraphics[width=0.95\columnwidth]{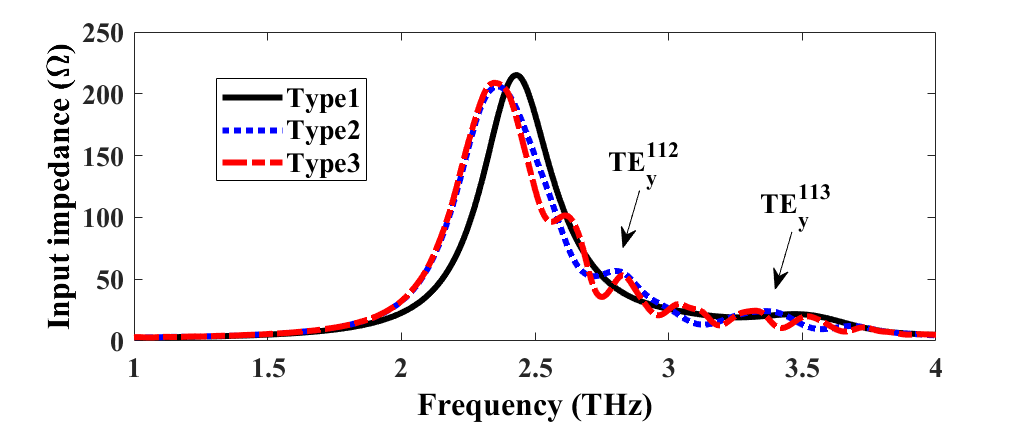}
\vspace{-0.2cm}
\caption{Real part of input impedance for the three antenna instances (see Table \ref{tab:comp}) with $\mu_c=0.8$ eV. Position of higher-order resonances are marked for Type 2.}
\label{fig:imp}
\vspace{-0.2cm}
\end{figure}  

\begin{figure*}[!tb] 
\centering
\subfigure[\label{fig:H1}2.4 THz (plasmonic and $TE_y^{111}$ modes)]{\includegraphics[width=0.282\textwidth]{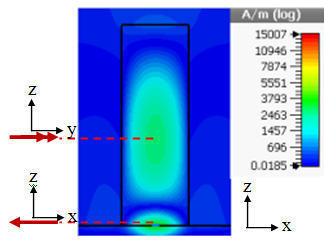} }
\subfigure[\label{fig:H2}2.8 THz ($TE_y^{112}$ mode)]{\includegraphics[width=0.28\textwidth]{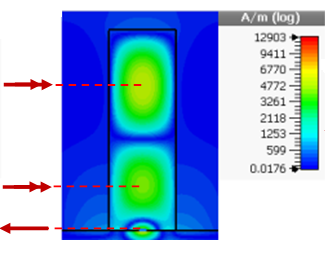} }
\subfigure[\label{fig:H3}3.3 THz ($TE_y^{113}$ mode)]{\includegraphics[width=0.268\textwidth]{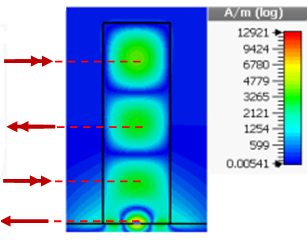} }
\vspace{-0.2cm}
\caption{H-fields of the Type 2 of the designed example antenna ($d_H = 60 \upmu$m) with $\mu_c=0.8$ eV, showing equivalent electric and magnetic dipoles at the different working points.}
\label{fig:H}
\vspace{-0.2cm}
\end{figure*}  

\vspace{0.1cm}
\noindent
\textbf{Evaluated Antennas:} In order to validate the proposed design, we assume the following structural parameters: $l=20$ $\upmu$m, $w=5$ $\upmu$m, $g=2$ $\upmu$m, ${\mu}_c=0.8$ eV, $d_{L1}=d_{L2}=100$ nm, $a=b=20$ $\upmu$m. Then, we consider three antennas with distinct DRA heights as summarized in Table \ref{tab:comp}. The heights are chosen so that different representative working points of the structure can be demonstrated. In the first instance (Type 1), $d_H$ ensures that the resonance frequency of dielectric resonator does not match with that of the graphene dipole. To this end, $d_H$ is small enough to consider that the radiation in the desired frequency is only caused by a pure plasmonic resonating mode (OC resonance). In the second instance (Type 2), the DRA is designed to resonate with the mode $TE_y^{111}$ at the OC resonance frequency of the graphene dipole. Consequently, the height of DRA is chosen equal to 60 $\upmu$m. In the third instance (Type 3), we seek to make the DRA resonate at a higher-order mode $TE_y^{112}$ to enhance the gain in the same frequency than the OC resonance of the graphene dipole. To this end, the height of DRA is selected equal to 120 $\upmu$m.

\vspace{0.1cm}
\noindent
\textbf{Methodology and results:} The three antennas are modeled and simulated with CST MWS. We report results obtained with the surface impedance modeling methodology and time domain solver. Although not shown for brevity, frequency solver and permittivity modeling yielded almost identical results.

Fig. \ref{fig:imp} illustrates the impedance matching performance. The input impedance of the antennas for the three considered case scenarios is calculated. It is observed that the OC resonance frequency of graphene dipole for Type 1 (pure plasmonic antenna) is near 2.5 THz while it is about 2.4 THz for Type 2 and 3 (hybrid antennas) because of the loading of the graphene dipole by the dielectric resonators. Hybrid antennas show impedance oscillations after 2.5 THz due to the presence of higher-order modes (e.g., 2.8 THz and 3.3 THz for Type 2).

Fig. \ref{fig:H} illustrates the simulated H-field distribution of Type 2 in three frequencies. According to Fig. \ref{fig:imp} and Fig. \ref{fig:H1}, $TE_y^{111}$ mode of DRA and OC resonance of graphene dipole occur simultaneously at 2.4 THz. $TE_y^{112}$ and $TE_y^{113}$ modes of type 2 resonate at 2.8 (Fig. \ref{fig:H2}) and 3.3 THz (Fig. \ref{fig:H3}), respectively. These points agree well with theoretical values from Eq. \eqref{eq:modes}. It should be noticed that the plasmonic mode on the graphene dipole radiates like an electric dipole along $x$ direction and the dielectric mode on DRA acts like a $z$-oriented array of magnetic dipoles along $y$ direction, whose number depends on the mode being excited within the DRA. These equivalent electric and magnetic dipoles are shown in Fig. \ref{fig:H}.

\begin{figure}[!t] 
\centering
\includegraphics[width=0.95\columnwidth]{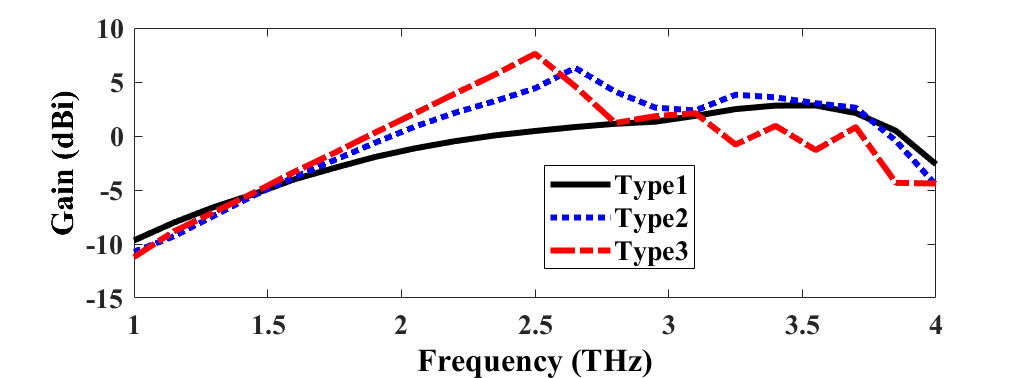}
\vspace{-0.2cm}
\caption{Gain as a function of frequency of the designed example antenna with $\mu_c=0.8$ eV at 2.4 THz for the three antenna instances.}
\label{fig:gain}
\vspace{-0.2cm}
\end{figure}  

The gain (IEEE) is evaluated as a function of the frequency. Fig. \ref{fig:gain} shows that the gain of graphene dipole loaded by DRAs (Type 2 and Type 3) is evidently enhanced compared with pure plasmonic antenna (Type 1). A gain of 4 dBi and 7 dBi is observed for the $TE_y^{111}$-mode DRA (Type 2) and $TE_y^{112}$-mode DRA (Type 3), respectively, whereas the radiation efficiency in both cases is around 70\% and achieved at around 2.4 THz. This represents a clear improvement in terms of antenna performance with respect to the pure plasmonic antenna, which shows a gain of 0.5 dBi and a radiation efficiency of 65\% at around 2.5 THz. Table I summarizes the results.

\begin{table}[!t] 
\caption{Summary of antenna types analyzed in this work.}
\vspace{-0.2cm}
\label{tab:comp}
\footnotesize
\centering
\begin{tabular}{|c|c|m{2.45cm}|c|c|} 
\hline
Structure & $d_H$ & Radiation mechanism & $e_{r}$ & Gain \\
\hline 
Type 1	& 5 $\upmu$m & Purely plasmonic & 65$\%$ & 0.5 dBi  \\ \hline
Type 2 	& 60 $\upmu$m & $TE_y^{111}$ DRA and plasmonic & 70$\%$ & 4 dBi \\ \hline
Type 3 	& 120 $\upmu$m & $TE_y^{112}$ DRA and plasmonic & 70$\%$ & 7	dBi \\ \hline
\end{tabular}
\vspace{-0.3cm}
\end{table}

\begin{figure*}[!t] 
\centering
\subfigure[\label{fig:pat1}Type 1 ($d_H = 5 \upmu$m)]{\includegraphics[width=0.25\textwidth]{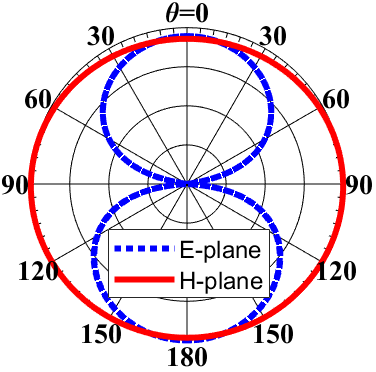} }
\subfigure[\label{fig:pat2}Type 2 ($d_H = 60 \upmu$m)]{\includegraphics[width=0.25\textwidth]{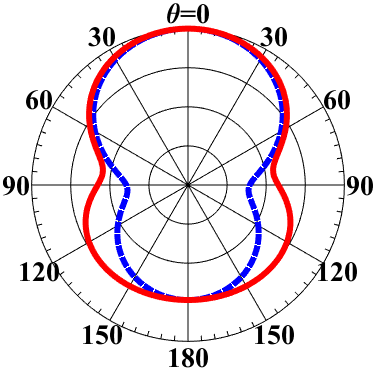} }
\subfigure[\label{fig:pat3}Type 3 ($d_H = 120 \upmu$m)]{\includegraphics[width=0.25\textwidth]{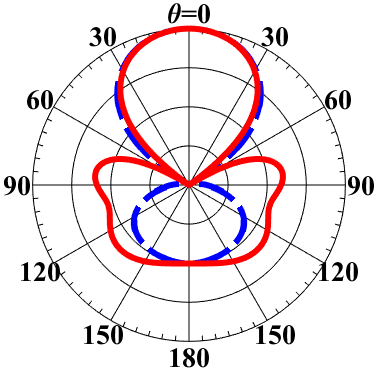} }
\vspace{-0.2cm}
\caption{Radiation patterns of the designed example antenna with $\mu_c=0.8$ eV at 2.4 THz for the three considered case scenarios.}
\label{pat}
\vspace{-0.3cm}
\end{figure*}  

Fig. \ref{pat} plots the radiation patterns of the three antennas at 2.4 THz. In Type 1, the dielectric modes do not resonate at the desired frequency and the radiation is only due to plasmonic dipole-like resonance, leading to a radiation similar to the hertzian dipole. It is also observed that the rectangular DRA operating in a higher order mode (Type 3) radiates a more directive pattern than the dominant mode (Type 2). The simple radiation modeling of the proposed antenna by means of an array of equivalent electric and magnetic dipoles can verify a significant increase in the directivity and gain.   

In summary, several points should be generally considered to drive the design of the proposed antenna. First, design a graphene plasmonic dipole operating in the OC resonance by using the approximate formula of Eq. \eqref{eq:spp} to fix the resonance frequency. Second, set the dimensions of the dielectric resonator by using the approximate formula of Eq. \eqref{eq:modes} taking into consideration the frequency and area limitations of the application. Finally, the performance of the antenna can be optimized with the help of a full-wave simulator.

\vspace{7pt}
\section{Conclusion}
\label{sec:conc}
In this paper, a novel THz antenna is proposed based on coupling of a dielectric wave mode and a plasmonic graphene mode. One-dimensional and two-dimensional analyses suggest the use of rather high chemical potential and H-layer thickness to have strong plasmonic--dielectric coupling. Results of the antenna as a three-dimensional structure show that an gain improvement of 6.5 dB is achieved by exciting the DRA at higher-order modes. The advantage of this approach for enhancing gain compared to other techniques such as the use of planar arrays is a much smaller area overhead, which is an important consideration for many applications where space limitations are of major concern.


%



\section*{Acknowledgment}

This work has been partially funded by Iran's National Elites Foundation (INEF), the Spanish Ministry of \emph{Economía y Competitividad} under grant PCIN-2015-012, the European Union under grant H2020-FETOPEN-736876, and the German Research Foundation (DFG) under grants HA 3022/9-1 and LE 2440/3-1.




\bibliographystyle{IEEEtran}
\bibliography{DRA-ref}
%

%








\end{document}